\documentclass[twocolumn,showpacs,amsmath,amssymb]{revtex4}
\usepackage{dcolumn}
\usepackage{graphicx}
\begin{document}
\title{Interplay between Josephson effect and magnetic interactions in double quantum dots}
\author{F. S. Bergeret}
\author{A. Levy Yeyati}
\author{A. Mart\'{i}n-Rodero}
\affiliation{Departamento de F\'{i}sica Te\'{o}rica de la Materia
Condensada C-V, Universidad Aut\'{o}noma de Madrid, E-28049
Madrid, Spain}
\begin{abstract}
We analyze the magnetic and transport properties of
a double quantum dot coupled to superconducting leads. In addition to the
possible phase transition to a $\pi$ state, already present in the single
dot case, this system exhibits a richer magnetic behavior due to the competition
between Kondo and inter-dot antiferromagnetic coupling. 
We obtain results for the Josephson current which may help to understand recent 
experiments on superconductor-metallofullerene dimer junctions. We show that 
in such a system the Josephson effect can be used to control its magnetic 
configuration. 
\end{abstract}
\pacs{74.50.+r,73.63.-b,75.20.Hr,73.21.La}
\maketitle
Quantum dot (QD) devices provide a unique 
opportunity to study the interplay between different
basic quantum phenomena. Thus, for instance, great advances in the 
understanding of Kondo physics have been achieved since the observation
of the Kondo effect in semiconducting quantum dots \cite{cronenwett}.
More recently double quantum dot (DQD) structures
have been proposed for studying the competition between
the Kondo effect and the inter-dot antiferromagnetic coupling 
\cite{N-DQD-N}.
An additional interesting ingredient is introduced
when these systems are connected to superconducting electrodes 
\cite{generic_sds}.
In this case the electron pairing in the leads appears as a competing
mechanism to both the Kondo and other type of magnetic interactions
that could be present.
For a single quantum dot placed between two
superconductors this competition can 
lead to a suppression of the Kondo effect and the appearance of an
unscreened magnetic moment, corresponding to a quantum phase
transition to the so-called $\pi$-state with a reversal of the
sign of the Josephson current \cite{pi,vecino}.
On the experimental side, great progress in the physical realization of
these systems is being achieved by  
structures consisting of nanotubes or fullerene molecules attached 
to metallic electrodes \cite{buitelaar,orsay}. 
In Ref. \cite{orsay} electron transport through
superconductor- metallofullerene molecules
(Gd@C$_{82}$)-superconductor junctions was analyzed. Strong
features associated with superconductivity were observed for the
case of junctions containing a molecular dimer. As pointed out in 
\cite{orsay}, the observed non-monotonic dependence of the low bias
current as a function of temperature could be related
to a change in the magnetic configuration of the Gd atoms.
DQD systems coupled to superconducting electrodes have been
theoretically analyzed in Refs. \cite{loss,zhu}. However, 
these works considered geometries and ranges of parameters
which do not correspond directly to the situation in the
experiments mentioned above.

In this Letter we provide an analysis, based on exact diagonalizations
and mean field slave boson techniques, of the
interplay between the Josephson effect, Kondo correlations and 
antiferromagnetic coupling in S-DQD-S systems. Like in the
single S-QD-S case we identify phases in which the sign of the
Josephson coupling is reversed. The situation in the S-DQD-S
system is however richer from the point of view of its magnetic
configuration. We show that when the system
is coupled to localized spins as in the experimental situation
of Ref. \cite{orsay} their relative orientation 
can be influenced by the Josephson current through the device.
We claim that these properties provide a way to control the magnetic 
configuration of such a nanoscale system.

The system depicted in Fig. 1 consists of two coupled quantum dots
in series, placed between two superconducting electrodes. The
electronic degrees of freedom are represented by a double Anderson
model with a single spin-degenerate level on each QD. The
corresponding Hamiltonian is given by
\begin{eqnarray}
\hat{H}_{el}&=& \hat{H}_L + \hat{H}_R + \sum_{i,\sigma}\epsilon_{i\sigma}\hat{n}_{i\sigma}
+U\sum_i\hat{n}_{i\uparrow}\hat{n}_{i\downarrow} \nonumber \\
&& +\hat{H}_{12}+\hat{H}_{1L}+\hat{H}_{2R} \, ,\label{hel}
\end{eqnarray}
where the index $i=1,2$ identify each QD; the terms $\hat{H}_L$ and
$\hat{H}_R$ describe the uncoupled leads as BCS superconductors characterized
by a complex order parameter $\Delta e^{i\phi_{L,R}}$.
$\hat{H}_{12}$ is  the coupling term between the dots given by $\hat{H}_{12}=
\sum_{\sigma} t_{12} \hat{c}^{\dagger}_{1\sigma}\hat{c}_{2\sigma} + \mbox{h.c.}$.
The last two terms correspond to the coupling between the dots and
the electrodes,
$\hat{H}_{1L(2R)}=\sum_{k\sigma}t_{L,R}\hat{c}^{\dagger}_{1,(2)\sigma}\hat{c}_{kL(R)\sigma}+{\rm
h.c.}$. The Coulomb interaction within
each dot is described by the $U$ term.
The coupling of the QD's with two localized magnetic moments,
which in the experiments of Ref. \cite{orsay} are provided by the
Gd impurities, can be modeled by an additional term in the
Hamiltonian of the form
\begin{equation}
\hat{H}_{int} = J \vec{S}_1.\vec{\sigma}_1 + 
J \vec{S}_2.\vec{\sigma}_2 \; , \label{hint}
\end{equation}
where $\vec{\sigma}_{1,2}$ are the electronic spin operators in the dots, while
$\vec{S}_{1,2}$ denote the localized spins.
We assume, in accordance to Refs. \cite{orsay,furukawa}, that
the magnetic coupling, $J$, is much weaker than the other energies involved in 
the problem,
which allows to introduce its effect as a perturbation in a second stage. 
There exists some controversy concerning the sign of $J$ in Gd@C82 although recent 
studies suggest that this coupling is antiferromagnetic \cite{furukawa}. The actual sign 
of $J$ is, however, not essential for the main effects discussed below. These studies also
suggest a large magnetic moment associated to the Gd impurities, which allows to 
consider $\vec{S}_{1,2}$ in our model as classical.
We will focus in the case of strong coupling between the
QD's by taking $t_{12}/\Delta=10$, which roughly 
corresponds to the estimates \cite{orsay2}
for the experimental situation of Ref. \cite{orsay}.
The results of this reference also suggest good coupling to the
leads ({\it i.e.} $t_{L,R} > \Delta$).   
\begin{figure}
 \includegraphics[scale=.3,angle=0]{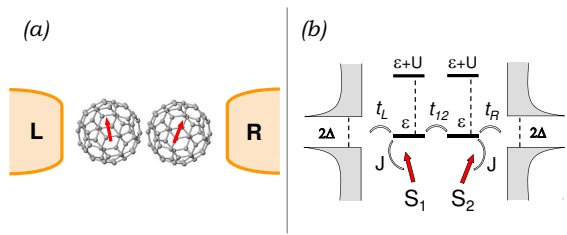}
\caption{(color online) Schematic view of the studied structure  (a) and
involved energies (b). This system should  model two fullerenes
doped with Gd atoms in contact with two superconducting reservoirs
(L, R).}
\end{figure}

A first insight into this problem can be provided by analyzing the 
$\hat{H}_{el}$ ground state properties as a function of the model 
parameters. For this purpose we rely on an approximation
consisting in taking the zero bandwidth limit (ZBWL) for the 
superconducting electrodes. The validity of this approach has 
been discussed for other superconducting junctions in Refs. \cite{vecino,affleck}. 
In this limit the Hilbert space of the S-DQD-S system is restricted 
to 4$^4$ states and $\hat{H}_{el}$ can be diagonalized exactly. 
In the superconducting case we
distinguish four different ground states: the pure $0$ and $\pi$ states
for which the energy as a function of the superconducting phase difference
$\phi=\phi_L - \phi_R$ has a minimum at $\phi=0$ and $\pi$ respectively; and 
two intermediate phases, which are designed as $0'$ and $\pi'$ depending of 
the relative stability of each minima \cite{vecino}. Fig. 2
illustrates the $(\epsilon,U)$ phase diagram for 
two different values of $t_R=t_L$: $2\Delta$ (panel (a)) and
$2.5\Delta$ (panel (b)). We show only the range of $\epsilon$ 
which corresponds to a charge per dot varying between 0 and 1, where
the transition to the $\pi$ state can take place \cite{senapati}. 
\begin{figure}
\includegraphics[scale=.4,angle=0]{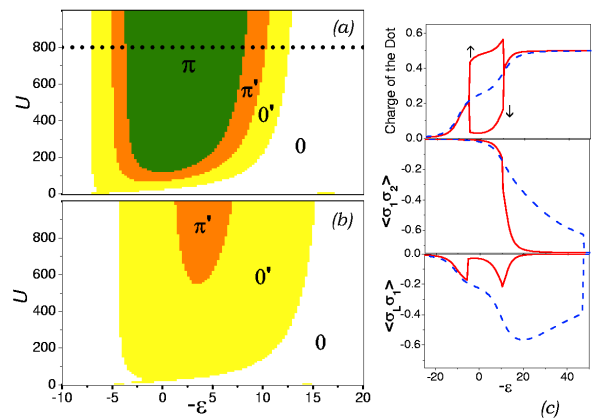}
\caption{(color online) $(U,\epsilon)$-phase diagram for $t_{12}=10\Delta$,
$t_L=t_R=2\Delta$ (a) and $t_L=t_R=2.5\Delta$ (b) indicating the
0, $0'$, $\pi'$ and $\pi$ regions.  On (c) we show the $\epsilon$
dependence along the line $U=800 \Delta$ indicated in (a)
of (from top to bottom): the occupation number for spin
up and spin down electrons; the interdot spin correlation
function, and between the DQD spin and the electrodes.
The dashed (solid) lines corresponds to the normal
(superconducting) state.}
\end{figure}
As shown in Fig. 2, for $t_L=t_R=2\Delta$ all phases
0, 0', $\pi'$ and $\pi$ appear at
the transition region. A more detailed understanding of the ground
state properties is provided by analyzing 
non-local spin correlation functions of the form
$<\vec{\sigma}_{\mu} \vec{\sigma}_{\nu}>$.
We choose  the line in the phase
diagram which corresponds to a large intradot Coulomb interaction, $U=800\Delta$, 
and show the evolution of these correlation functions with $\epsilon$ in Fig. 2 (c) 
together with the occupation numbers $n_{\uparrow,\downarrow}$ for 
each dot. 
The appearance of a 1/2 magnetic moment for
the full S-DQD-S system is signaled by the broken symmetry
$n_\downarrow\neq n_\uparrow$. The function 
$<\vec{\sigma}_1\vec{\sigma}_2>$ measures the correlation between the
electron spins in the two dots. As can be seen in the middle panel 
of Fig. 2 (c) it evolves continuously
from $0$ to $-3/4$, the latter value corresponding to a complete
antiferromagnetic (AF) correlation. It is worth noticing that this
AF tendency is more pronounced in the superconducting case as the
presence of the superconducting gap reduces the
number of low energy excitations capable to screen the spin 
in the dot region, {\it i.e.} it leads to a partial
suppression of the Kondo correlations. 
In the present range of parameters, with a strong interdot hopping, the
Kondo regime corresponds to the formation of a spin singlet between 
an electron in the bonding state of the DQD "molecule" 
and the electrons in the leads. 
This correlation is reflected in the behavior of  
$<\vec{\sigma}_{L(R)} \vec{\sigma}_{1(2)}>$, which becomes
increasingly negative in the Kondo regime.
As can be observed in the lower panel of 
Fig. 2 (c), 
Kondo correlations are strongly suppressed by superconductivity
compared to the normal case. In fact, it is
in the range of parameters corresponding to the Kondo region
in the normal state where the $\pi$ state appears. The normal state
itself exhibits a transition from the Kondo to the AF regime for
$-\epsilon \gg t_{12}$, signaled by
$<\vec{\sigma}_{L(R)}\vec{\sigma}_{1(2)}>\rightarrow 0$ and
$<\vec{\sigma}_1\vec{\sigma}_2>\rightarrow -3/4$. This transition
roughly corresponds to the situation where the AF coupling between the
dots $\sim t_{12}^2/U$ becomes larger than the AF coupling of 
the dots with the leads $\sim 2t_{L,R}^2/|\epsilon-t_{12}|$. 
If the hopping to the leads is increased, the suppression of the Kondo effect by
superconductivity becomes less effective. As shown in Fig. 2 (b), 
for $t_L = t_R = 2.5 \Delta$ the system exhibits
only three phases (0, 0' and $\pi'$) within the range of parameters 
considered. 
Further increase of $t_{L,R}$  would lead
to a complete suppression of the $\pi'$ and $0'$ phases.

In order to go beyond the ZBWL approximation and include the 
finite bandwidth of the electrodes we use an appropriate
slave-boson representation of Hamiltonian (\ref{hel}).
In Ref. \cite{zaikin} the $U\rightarrow\infty$
mean field slave-boson approach \cite{coleman} was used to study the 
single QD system with superconducting electrodes. However, in order to
describe the main features of these systems including the possibility of 
unscreened magnetic moments it is necessary to use the more general
representation of Ref. \cite{ruckenstein}, which is valid for
finite values of $U$ and allows for possible magnetic solutions
\cite{dong}.
Following Ref. \cite{ruckenstein} the auxiliary
Bose fields are designed  by $\hat{e}_i$ (empty state), $\hat{p}_{i\alpha}$
(single occupied state corresponding to spin $\sigma$) and $\hat{d}_i$
(double occupied state) and we define the operator
$\hat{z}_{i\sigma}=(1-\hat{d}_i^2-\hat{p}_{i\sigma}^2)^{-1/2}
(\hat{e}_i\hat{p}_{i\sigma}+\hat{p}_{i\bar{\sigma}}\hat{d}_i)
(1-\hat{e}_i^2-\hat{p}_{i\bar{\sigma}}^2)^{-1/2}$, 
where $i=1,2$ denotes the two different
QDs. In the enlarged space the Hamiltonian (\ref{hel}) has the
form
\begin{widetext}
\begin{eqnarray}
\hat{H}_{el} &=& \hat{H}_L+\hat{H}_R+
\sum_{i\sigma}\epsilon_i \hat{f}_{i\sigma}^{\dagger}\hat{f}_{i\sigma}+
\sum_iU\hat{d}_i^{\dagger}\hat{d}_i +\sum_{\sigma}
t_{12}(\hat{z}_{1\sigma}^{\dagger}\hat{z}_{2\sigma}
\hat{f}_{1\sigma}^{\dagger}\hat{f}_{2\sigma}+{\rm h.c.} + 
\sum_{k,\sigma}t_{L(R)}(\hat{z}_{1(2)\sigma}^{\dagger}
\hat{f}_{1(2)\sigma}^{\dagger}\hat{c}_{kL(R)\sigma}+h.c.)\nonumber\\
&&-\sum_i\alpha_i(\hat{e}_i^{\dagger}\hat{e}_i+\hat{d}^{\dagger}\hat{d}_i+\sum_{\sigma}\hat{p}_{i\sigma}^{\dagger}\hat{p}_{i\sigma}-1)
-\sum_{i\sigma}\beta_{i\sigma}(\hat{f}_{i\sigma}^{\dagger}\hat{f}_{i\sigma}
-\hat{p}_{i\sigma}^{\dagger}\hat{p}_{i\sigma}-\hat{d}^{\dagger}\hat{d}_i)\label{sbham}
\end{eqnarray}
\end{widetext}
where the $\hat{f}_{i\sigma}$ are fermionic operators and $\alpha_i$,
$\beta_{i\sigma}$ are the Lagrange multipliers corresponding to
the constraints
$\hat{e}_i^{\dagger}\hat{e}_i+\hat{d}^{\dagger}\hat{d}_i+
\sum_{\sigma}\hat{p}_{i\sigma}^{\dagger}\hat{p}_{i\sigma}=1$
and $\hat{f}_{i\sigma}^{\dagger}\hat{f}_{i\sigma}=
\hat{p}_{i\sigma}^{\dagger}\hat{p}_{i\sigma}+\hat{d}^{\dagger}\hat{d}_i$. 
The particular definition of the $\hat{z}_{i\sigma}$ 
operators warrants that the exact solution in the $U\rightarrow 0$ limit 
is recovered \cite{ruckenstein}. 
Within a mean field approximation we 
replace the Bose operators in Eq. (\ref{sbham}) by their mean 
values $e_i$, $d_i$, $p_{i\sigma}$ and $z_{i\sigma}$. In this
approximation the Hamiltonian parameters are renormalized
according to $\tilde{\epsilon}_{\sigma i}= \epsilon_i-\beta_{i\sigma}$,
$\tilde{t}_{12 \sigma}=t_{12} z_{1\sigma}z_{2\sigma} $ and
$\tilde{t}_{L(R)\sigma}=t_{L(R)}z_{1(2)\sigma}$.  The mean
values of the Bose operators must be determined self-consistently
by minimizing the effective action \cite{ruckenstein}.

Both the current through the dots $I = ie/\hbar \sum_{\sigma} 
\tilde{t}_{12 \sigma} [<\hat{f}^{\dagger}_{1\sigma} \hat{f}_{2\sigma}>
- <\hat{f}^{\dagger}_{2\sigma} \hat{f}_{1\sigma}>]$ and their
occupations $n_{i\sigma} = <\hat{f}^{\dagger}_{\sigma} \hat{f}_{i\sigma}>$
can be calculated using standard Green function techniques \cite{alvaro}.
We have solved numerically the mean field equations and computed
the Josephson current through the DQD system for a certain set of
parameters. In Fig. 3 we show the current-phase relation for
$U=800\Delta$,  $t_{12}=10\Delta$ and two different values of the parameter
$\Gamma_{L,R}=\pi t_{L,R}^2\rho_{L,R}(E_F)$,
where $\rho_{L,R}(E_F)$ is the electrodes normal density of states
at the Fermi energy. We see that for
$\Gamma_L=\Gamma_R=2.25 \Delta$ the system evolves from the
$0$ to the $\pi$ state as $\epsilon$ varies between $\sim -5\Delta$ and
$\sim -2\Delta$ going back to the 0 state for $\epsilon \sim 7\Delta$.
In the case of a larger coupling, $\Gamma_{L,R}=4 \Delta$,  
the pure $\pi$ state is never reached, in good qualitatively 
agreement with the results obtained within the ZBWL.
It is worth noticing that the occurrence of the $\pi$
state requires $\Delta$ to be larger than an energy scale $\sim
\sqrt{\Gamma U/2}\exp{(-\pi|\epsilon-t_{12}|/2\Gamma)}$, associated 
with the Kondo effect of the singly occupied bonding level, which for 
$t_{12}/\Gamma > 1$ can be much smaller than the effective Kondo 
temperature estimated for the normal DQD system \cite{aguado}.
\begin{figure}
\includegraphics[scale=0.3,angle=0]{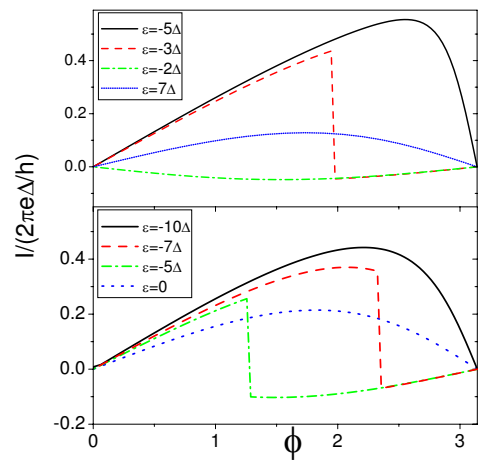}
\caption{(color online) The current-phase relation for $t_{12}=10\Delta$,
$U=800\Delta$, $\Gamma_L=\Gamma_R=2.25 \Delta$ (upper panel) and
$\Gamma_L=\Gamma_R=4 \Delta$ (lower panel) }
\end{figure}

Let us now analyze how the behavior of the electronic system
could influence the configuration of the localized spins 
by means of their magnetic coupling given by Eq. (\ref{hint}). 
One would expect that
the appearance the magnetic $\pi$-state could give rise to a
change in this configuration with respect to the case of normal 
electrodes. In fact, for small $J$ the total
energy can be expanded as
\begin{equation}
E(h_1,h_2) \simeq E(0,0) + \sum_{\mu=1,2} a_{\mu} h_{\mu} + \sum_{\mu,\nu} a_{\mu,\nu}
h_{\mu} h_{\nu} ; 
\end{equation}
where $h_{\mu} = J S_{\mu,z}$, $S_{\mu,z}$ being the $z$-component of the
localized spin.
Notice that in general, for non-magnetic situations, only the quadratic
correction appears as in the well known RKKY interaction \cite{rkky}. For the range
of parameters where the $\pi$ states appear this correction would be positive 
leading to a AF configuration of the localized spins. However,
the broken symmetry, $n_{\uparrow} \neq n_{\downarrow}$, in the $\pi$ state gives rise
to non-vanishing linear corrections which favor the parallel (F) configuration.
This is illustrated by the insets in the upper panel of Fig. 4 which show the
behavior of the total energy as a function of $h=|h_1|=|h_2|$ at $\phi=0$ and
$\phi=\pi$ respectively in the 0' region of the phase diagram. A simple image 
of this effect is that the unscreened magnetic moment appearing in the 
$\pi$-state acts as a local magnetic field which tends to align the localized spins.
The full phase dependence of the total energy for finite $J$ in the F and 
the AF configurations is depicted in the upper panel of Fig. 4. 
As can be observed the range of stability of the $\pi$ state is increased in the
F configuration, which in turn has a noticeable effect in the Josephson current
(lower panel of Fig. 4). Thus, our results predict that the system 
switches from the AF to the F configuration as the superconducting phase is 
sweeped from 0 to $\pi$. They also suggest that in the transition region the
F configuration is a metastable solution which could give rise to an hysteretic
behavior as a function of the superconducting phase difference.  
Although a direct comparison with the 
data of Ref. \cite{orsay} is not possible since the experiment was
performed  under non-equilibrium conditions, our results tend to support
that the non-monotonic behavior of the low bias current as a function of
temperature may be indeed related to a change of the magnetic
configuration of the Gd atoms. A more direct test of our predictions would
require the measurement of the supercurrent in a phase biased situation. \begin{figure}
\includegraphics[scale=.3,angle=0]{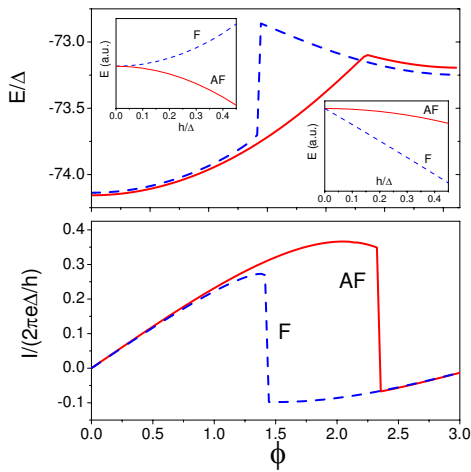}
\caption{(color online) Energy (upper panel) and current-phase relation 
(lower panel) corresponding to
the parallel (F, dashed line) and antiparallel (AF, full line) 
configuration of the localized spins. We have chosen $U=800\Delta$, 
$t_{12}=10\Delta$, $|h_1|=|h_2|=0.25\Delta$ and $\Gamma_L=\Gamma_R=2.25\Delta$. 
The insets show the behavior of the energy as a function of $|h_{1,2}|$
for the F and AF configuration at $\phi=0$ and $\phi=\pi$.}
\end{figure}

In conclusion we have studied the magnetic and superconducting
properties of a S-DQD-S system. We have shown that it can exhibit
a quantum phase transition to a $\pi$ state with an unscreened
1/2 magnetic moment in the dots region. 
When the system is coupled to localized spins as in the experimental
situation of Ref. \cite{orsay} a transition from an AF to a F configuration
can be induced by tunning the superconducting phase difference. 
These properties illustrate the possibility of controlling the magnetic 
configuration at the nanoscale by means of the Josephson effect.

{\it Acknowledgments:} The authors acknowledge EU for financial support
through the DIENOW network and Spanish CYCIT under contract FIS2005-06255.
Fruitful comments by H.Bouchiat and S. Gueron are also acknowledged.
FSB thanks the Oxford Theoretical Physics Department for hospitality during 
his stay.

\end{document}